\documentclass[onecolumn, usenatbib]{mnras}
\usepackage{times,graphicx,amssymb, amsmath, natbib}
\usepackage[T1]{fontenc}

\usepackage{epsfig}

\usepackage{amsmath}
\usepackage{amssymb}
\usepackage{epsfig}

\newcommand{\beq}{\begin{equation}}
\newcommand{\eeq}{\end{equation}}

\newcommand{\ber}{\begin{eqnarray}}
\newcommand{\eer}{\end{eqnarray}}

\def\apj{{Astroph.\@ J.\ }}
\def\mnras{{Mon.\@ Not.\@ Roy.\@ Ast.\@ Soc.\ }}

\def\aj{{Astron.\@ J.\ }}

\def\prd{{Phys.\@ Rev.\@ D\ }}

\def\nat{{Nature\ }}

\def\beq{\begin{equation}}
\def\eeq{\end{equation}}

\def\ber{\begin{eqnarray}}
\def\eer{\end{eqnarray}}

\begin{document}

\title{Cosmology and Astrophysics Using the Post-reionization HI}

\author[Guha Sarkar and Sen.]
{Tapomoy Guha~Sarkar $^{1}$, 
Anjan A Sen $^{2}$\\
$^{1}$Department of Physics, Birla Institute of Technology and Science, Pilani, Rajasthan, 333031. India, E-mail: tapomoy1@gmail.com.\\
$^{2}$Centre for Theoretical Physics, Jamia Millia Islamia, New Delhi-110025, India, E-mail:aasen@jmi.ac.in.
}

\maketitle
\date{\today}
\thispagestyle{empty}

\begin{abstract}
We discuss the prospects of using the redshifted 21~cm emission from neutral hydrogen in the post-reionization epoch to study our universe. The main aim of the article is to highlight the efforts of Indian scientists in this area with the SKA in mind. It turns out that the intensity mapping surveys from SKA can be instrumental in obtaining tighter constraints on the dark energy models. Cross-correlation of the HI intensity maps with the Ly$\alpha$ forest data can also be useful in measuring the BAO scale.

\end{abstract}

\maketitle

\section{Introduction}

The post-reionization ($z \lesssim 6$) evolution of neutral hydrogen (HI) is of great interest from the point of view of both astrophysics as well as cosmology. After reionization, the bulk of HI is expected to reside in galaxies in the form of clumpy, self-shielded clouds \citep{1986ApJS...61..249W,1991ApJS...77....1L,2000ApJ...543..552S,1997ApJ...484...31G,2005ApJ...635..123P}. 
%This picture is supported by observations of damped Lyman-alpha systems associated with FeII and MgII absorption systems, stacked observations of HI line emission from star forming galaxies, and direct observations of HI emission in individual HI-rich galaxies in the local Universe. 
The radio observations of the redshifted 21~cm line promise to be a powerful probe of HI, and it is expected that present and upcoming facilities like the GMRT\footnote{http://gmrt.ncra.tifr.res.in/}, MeerKAT\footnote{http://www.ska.ac.za/meerkat/}, CHIME\footnote{http://chime.phas.ubc.ca} and SKA\footnote{https://www.skatelescope.org} would play important roles in this area.

The 21~cm line can used in various ways to study the HI. It is already being used extensively in detecting and studying absorption systems (e.g., DLAs) along lines of sight towards distant radio sources. These studies are complementary to those done using UV / optical telescopes \citep{2009ApJ...696.1543P,2005ApJ...635..123P,2013A&A...556A.141Z}. An alternate way is to detect the HI in emission. The most standard method is to detect individual galaxies in HI through the so called HI galaxy redshift surveys \citep{2001MNRAS.322..486B,2004MNRAS.350.1195M,2005MNRAS.359L..30Z,Jaffe:2012ax,2013MNRAS.435.2693R,2005AJ....130.2598G,2010ApJ...723.1359M,2010MNRAS.403..683C,2007MNRAS.376.1357L,2009MNRAS.399.1447L}. Another way is to detect only the integrated HI emission of galaxies at each sky location without attempting to resolve the individual objects, a technique known as HI intensity mapping \citep{2001JApA...22..293B,2010Natur.466..463C,2013ApJ...763L..20M,2009MNRAS.399.1447L}. There has been recent surge of investigations aimed towards understanding dark energy using future  SKA observations \citep{TGS15, 2016arXiv160302087H,2015ApJ...803...21B,2015aska.confE.165Z, 2015aska.confE..19S, 2015aska.confE..24B}.

This article focuses mainly on Indian interests related to studying astrophysics and cosmology using HI emission in the post-reionization universe. The main science cases which have been considered here are (i) the constraints on dark energy using HI intensity mapping experiments, (ii) measuring the BAO using the cross correlation of intensity mapping and Ly$\alpha$ forest. 

\section{Dark energy constraints from intensity mapping experiments}

The large-scale clustering of the HI in the post-reionization epoch is expected to directly probe the nature of dark energy through the imprints of a given model on the background evolution and growth of structures. As a direct probe of cosmological structure formation, 21-cm intensity mapping may allow us to distinguish between dark energy models which are otherwise degenerate at the level of their prediction of background evolution. In the of context of SKA, this has been recently investigated by a group of Indian scientists including the present authors \citep{2016arXiv160302087H}.

\subsection{Constraints from angular power spectra}

To begin with, we consider a very general picture where the two dark components in the universe, dark matter (DM) and dark energy (DE),  are interacting with each other. For the dark energy part, we assume the simplest scenario of a canonical scalar field rolling slowly over a very flat potential. The visible sector of the matter consisting baryons is not coupled with DE. This scenario, also known as ``{\it Coupled Quintessence}" \citep{2000PhRvD..62d3511A,2004PhRvD..69j3524A} is a well studied phenemenological model for dark energy. One can easily switch off the interaction term between DM and DE to consider the case of minimally coupled scalar field dark energy models. The equations governing the spatially flat background universe are as follows \citep{2016arXiv160302087H}:

\begin{eqnarray}
 \ddot{\phi}+\frac{dV}{d\phi}+3H\dot{\phi}=C(\phi)\rho_{d} \nonumber \\ 
 \dot{\rho}_{d}+3H(\rho_{d})=-C(\phi)\rho_{d}\dot{\phi} \\
 \dot{\rho_b}+3H(\rho_b)=0 \nonumber\\
  H^2=\frac{\kappa^2}{3}(\rho_b+\rho_{d}+\rho_\phi). \nonumber \\
\end{eqnarray}

\noindent
Here $C(\phi)$ represents the coupling  between DM and DE. Subscript ``d" represents the DM sector and subscript ``b" represents the baryonic sector and an overdot represent differential w.r.t comoving time. We still do not know exactly the details physics of interaction bewteen DM and DE; hence our approach is phenomenological where we assume the function $C(\phi)$ to be a constant as in the previous work by Amendola and others \citep{2000PhRvD..62d3511A,2004PhRvD..69j3524A}. For $C=0$, we recover the minimally coupled scalar field dark energy model. Hence we can study both the coupled and uncoupled case in the same set up.

We study the growth of matter fluctuations in the linear regime and sub-horizon scales where we can safely igonore the dark energy perturbation. In this regime, Newtonian treament for density perturbation is sufficient for our purpose. With these assumptions, we can write the equations governing the growth of linear fluctuations in dark matter density as:

\begin{equation}
 \delta''_{d}+\left(1+\frac{\mathcal{H}'}{\mathcal{H}}-2 W x\right)\delta'_{d}-\frac{3}{2}(\gamma_{d}\delta_{d}
 \Omega_{d})=0.
\end{equation}

\noindent
Here prime denotes differentiation w.r.t to $\log a$ and  $x = \frac{\kappa \dot{\phi}}{\sqrt{6}H}$. Here $W = \frac{C}{\kappa}$, $\gamma_{d}=1+2 W^2$, ${\mathcal{H}}$ is the conformal Hubble parameter ${\mathcal{H}=aH}$ and $\delta_{d}$ is the linear density contrast for the DM. We solve the equation with the initial conditions at decoupling ($a \sim 10^{-3}$) when $\delta_{d} \sim a$ and $\frac{d\delta_{d}}{da} = 1$. 

\noindent
Taking the Fourier transform of the above equation, we define the linear growth function $D_{d}$ and the linear growth rate $f_{d}$ as
\begin{eqnarray}
\delta_{d k} (a) &\equiv& D_{d}(a)\delta_{d k}^{ini}\\
f_{d} &=& \frac{d \ln{D_{d}}}{d \ln{a}}.
\end{eqnarray}
The linear dark matter power spectrum given by
\begin{equation}
P(k,z) = A_0 k^{n_s} T^2(k) D_{d n}^2(z).
\end{equation}

\noindent
Here $A_0$ is the normalization constant and we fix it using $\sigma_{8}$ value at present from Planck 2015 observations \citep{Ade:2015xua}, $n_{s}$ is spectral index for the primordial density fluctuations generated through inflation, $D_{d n} (z)$ is growth function normalized such as it is equal to unity at $z=0$ i.e. $D_{d n}(z)=\frac{D_{d}(z)}{D_{d}(0)}$. $T(k)$ is the transfer function as given by Eisenstein and Hu \citep{EHU}.

To constrain dark energy behaviour using the 21-cm intensity mapping, the quantity of interest is the fluctuations in the HI 21-cm brightness temperature $\delta T_{b}$ and its corresponding angular power spectra which in the flat sky limit is given by \citep{datta1} 
\begin{equation} 
C_{l} = \frac{{\bar T}^2 \bar x_{HI}^2 b_T^2}{\pi r^2}\int
\limits_0^\infty dk_{\parallel} ~( 1 + \beta_T \mu^2)^2~ P(k,
z).
\end{equation}

\begin{figure*}
\begin{center} 
\resizebox{200pt}{160pt}{\includegraphics{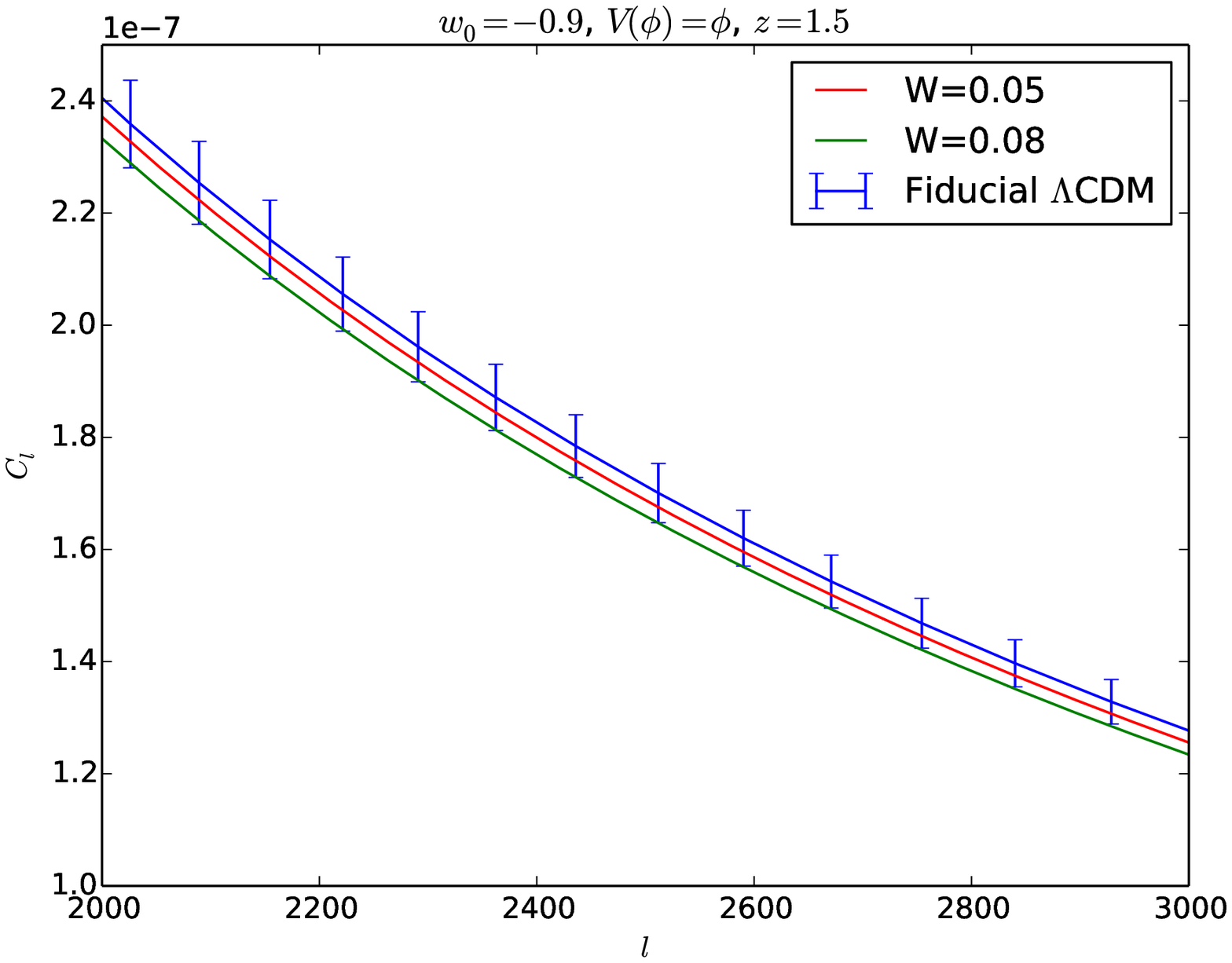}}
\hspace{1mm} \resizebox{200pt}{160pt}{\includegraphics{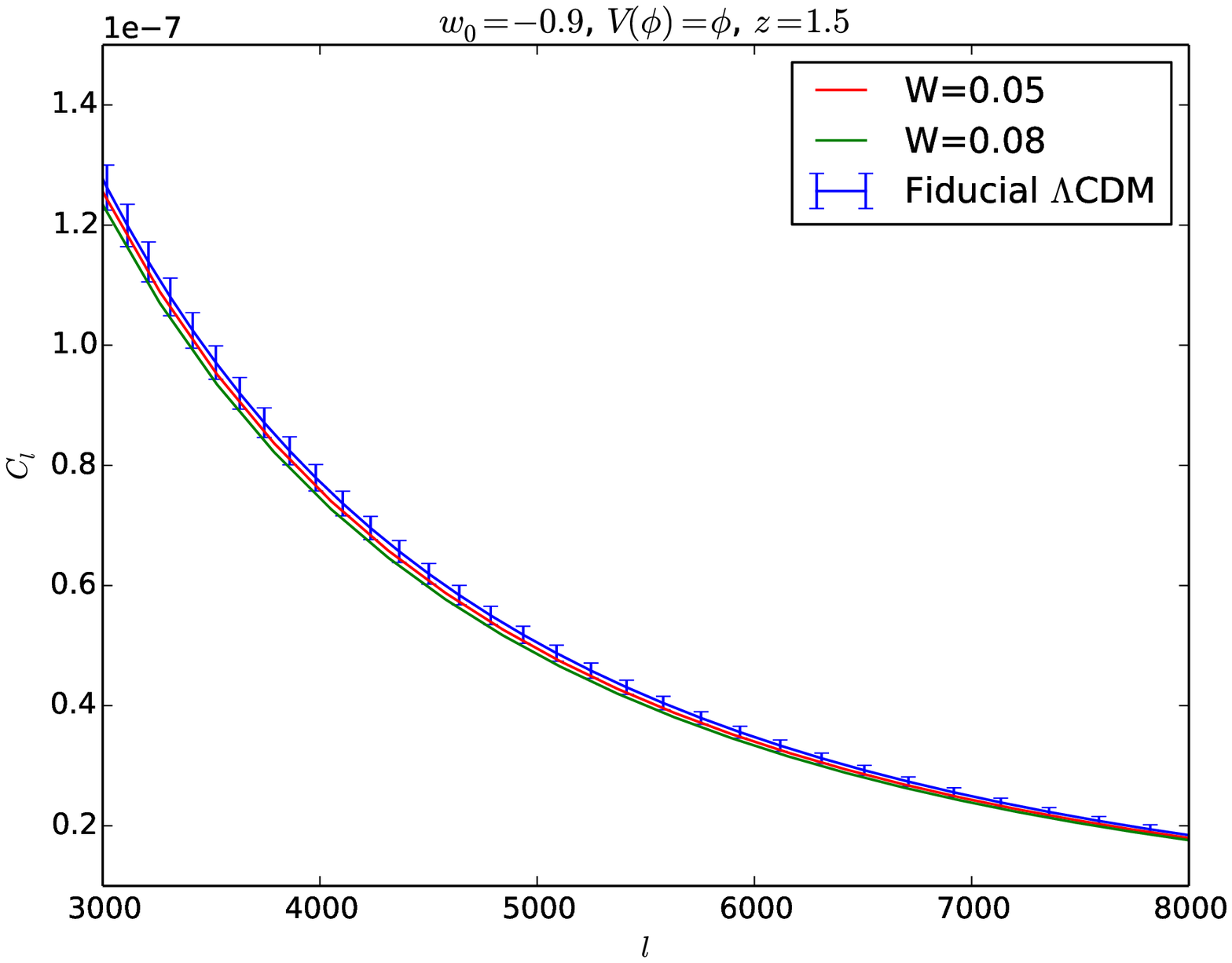}}\\
\resizebox{200pt}{160pt}{\includegraphics{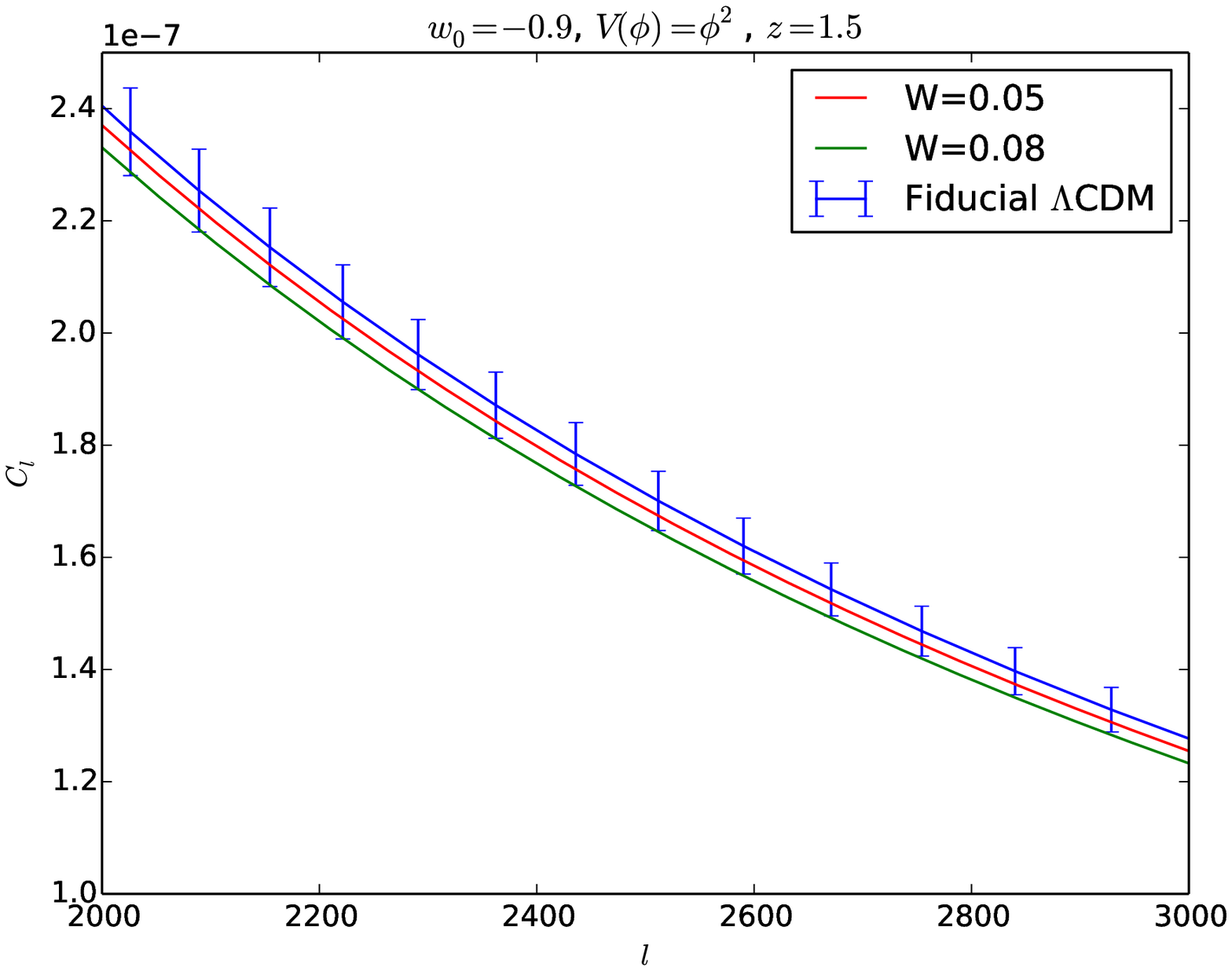}}
\hspace{1mm} \resizebox{200pt}{160pt}{\includegraphics{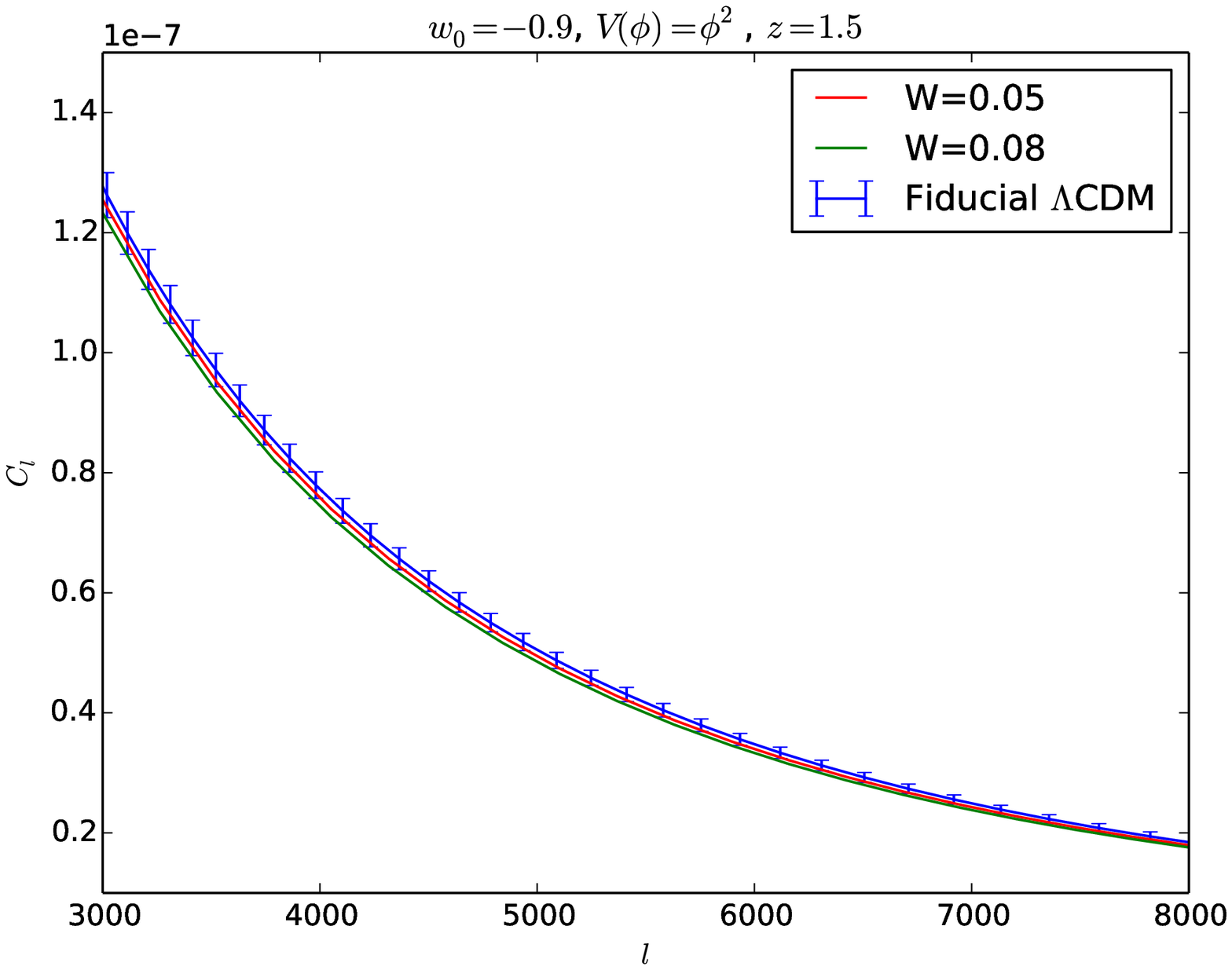}}\\
\resizebox{200pt}{160pt}{\includegraphics{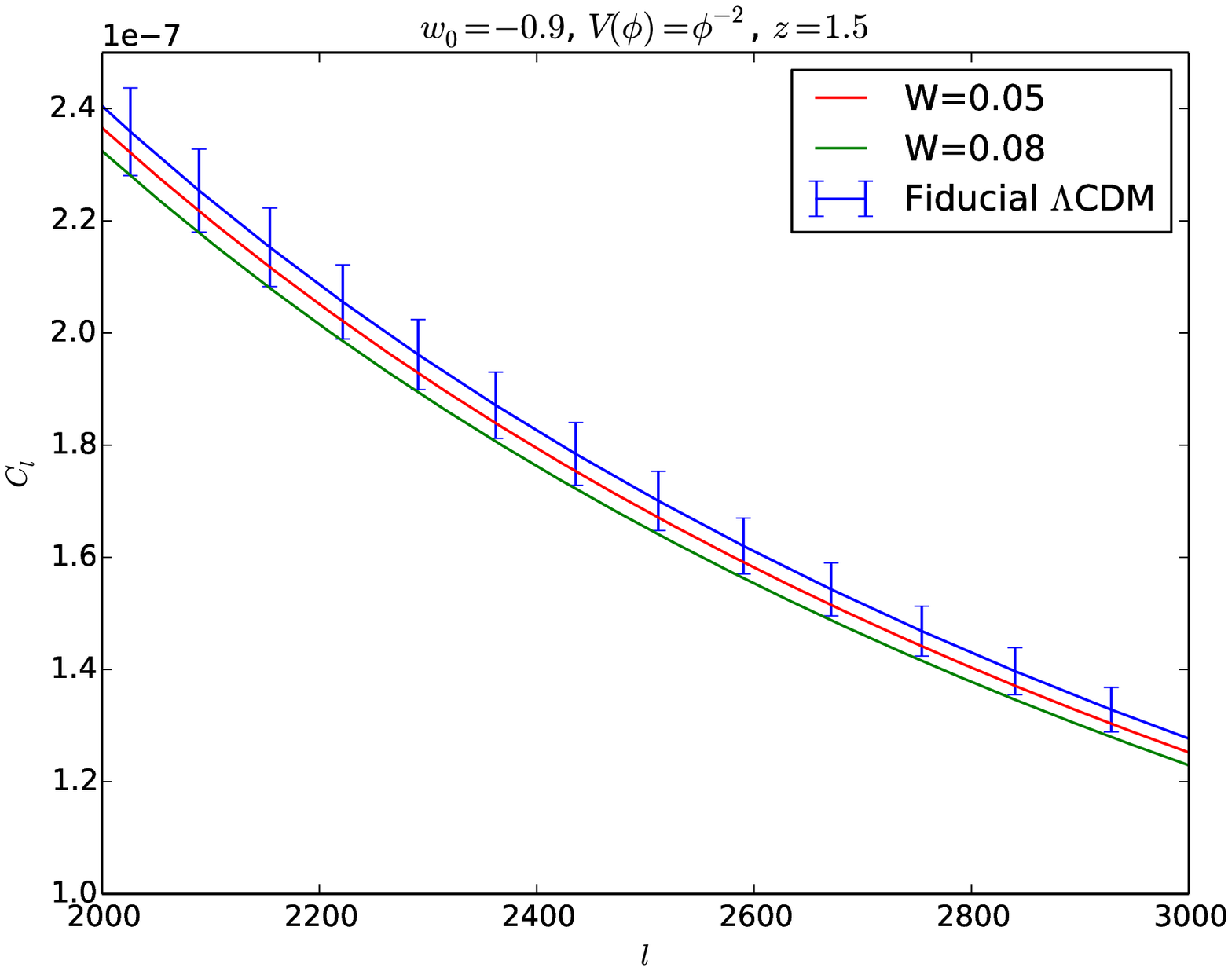}}
\hspace{1mm} \resizebox{200pt}{160pt}{\includegraphics{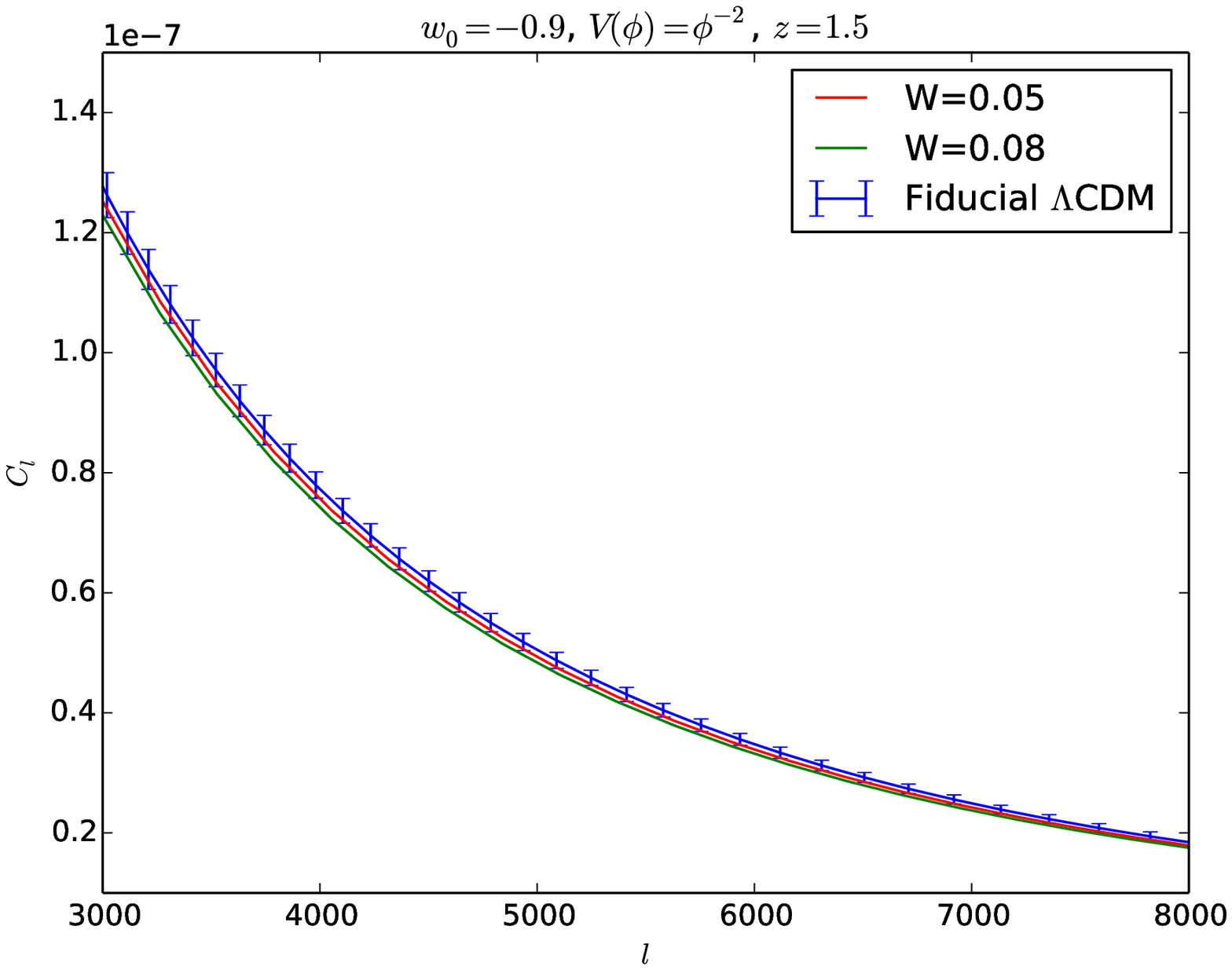}}\\

\end{center}
\caption{Angular power spectra for the fluctuations in HI 21-cm brightness tempertaure for different dark scalar field with different potentials $V(\phi)$ and coupling parameter $W$.  The left colomn is for $2000\leq l \leq 3000$ and the right colomn is for $3000\leq l \leq 8000$. }
\end{figure*}

\noindent
Here $P(k, z)$ is the dark matter power spectrum defined in equation (6). $k = \sqrt{k_{\parallel}^2 + \frac{l^2}{r^2}}$, $r$ being the comoving distance at a certain redshift $z$, $\mu =
\frac{k_{\parallel}}{k}$ and $\beta_T = \frac{f_{d}(z)}{b_T}$ where $f_{d}(z)$
is the growth factor for dark matter defined in equation (5). $b_{T}$ is the linear bias parameter. The average brightness temperature $\bar T$ at redshift $z$ is given by 
\begin{equation}
\bar T(z)=4.0 {\rm mK} (1+z)^{2}\left(\frac{\Omega_{b0}h^{2}}{0.02}\right)\left(\frac{0.7}{h}\right) \left( \frac{H_{0}}{H(z)}\right).
\end{equation}
\noindent
We assume $\Omega_{HI} = 10^{-3}$ for $ z<3.5$. This gives $\bar x_{HI} = 2.45\times 10^{-2}$  and we assumed this to be constant across the redshift range of our interest. Also our assumption of flat sky approximation is valid for $l > 200$ \citep{datta1}.

We consider a radio interferometric measurement of the power spectrum of the HI 21-cm brightness temperature. The noise in our measurement of angular power spectrum comes from instrumental noises on smalls scales and cosmic variances on large scales. This is given by
\begin{equation}
{\Delta C_l} = \sqrt{\frac{2}{(2l + 1)\Delta l f_{sky} N_p}} \left ( C_l + N_l \right ).
\end{equation}
where $N_p$ being the number of pointings of the radio interferometer, $f_{sky}$ is the fraction of sky observed in a single pointing, and $\Delta l $ is the width of the $l$ bin and $N_{l}$ is the noise power spectrum.

\noindent
To calculate $\Delta C_l$, we use the following specifications of SKA1-mid \footnote{http://www.skatelescope.org/wp-content/uploads/2012/07/SKA-TEL-SKO-DD-001-1\_BaselineDesign1.pdf}:

\noindent
We consider the fiducial redshift $z= 1.5$ which corresponds to an observing frequency of $568{\rm MHz}$ that falls in the
band of frequencies to be probed by SKA1-mid. The frequency bandwidth is $32 {\rm MHz}$ around the central frequency. The array consists of 200 dishes each of diameter $15$m. The
antennae are distributed in a manner such that $75\%$ of the dishes
are within $2.5$Km radius and the number density of antennae are assumed to
fall off radially as $ r^{-2}$.
 We note that  $ T_{sky} = 180\left (\frac{\nu}{180{\rm MHz}} \right) ^{-2.6}$ K. in our error estimates. However, the total system temperature $T_{syst}$ has a contribution from both the sky temperature and instrumental noise. In this  paper we have used  $  T_{syst} = 40 $K from  the SKA document \footnote{
https://www.skatelescope.org/uploaded/21705-130-Memo-Dewdney.pdf}. Our fiducial model for error calculations is a $\Lambda$CDM model with $\Omega_{\Lambda} = 0.7$, $\Omega_{b0} = 0.05$, $n_{s} = 1$, $h=0.7$ and $\sigma_{8} = 0.8$.
We also fix the value of the  constant linear bias at the value $1.0 $. For detail derivation of the noise, we refer readers to \citep{2016arXiv160302087H}. The noise is estimated by considering a radio observation of 10,000 hrs distributed equally over 40 pointings. It has been noted that in intensity mapping experiments it is imperative to cover large observational volumes rather than have a deep probe of a small part of the sky.

The results of our analysis is shown in figures 1 and 2. In figure 1, we show the angular power spectra $C_{l}$ for the fluctuations in the HI 21-cm brightness temperature for scalar field model with different potentials and for different values for the coupling parameter $W$. We normalize all the models to have the same present day value of the equation of state, $w_{0} = -0.9$ and density paremeter for the dark energy $\Omega_{\phi 0}= 0.7$. In each plot, we also show the $C_{l}$ for our fidcucial $\Lambda$CDM model and the corresponding error bars for SKA1-mid. The figures show that for higher values of the coupling parameter $W$, it may be possible to distinguish scalar field models from $\Lambda$CDM by SKA1-mid for smaller angular scales (higher $l$'s). We concentrate scales in the range of $ l  \geq 2000 $ where we neglect the relativistic corrections as well as corrections due to DE perturbations.

Next to see what happens when we turn off the coupling between the DM and DE, in figure 2 (left), we show the same plot as in figure 1 but only for linear potential with $W = 0.8$ and $W =  0$ which clearly shows that it may not be possible to distinguish minimally coupled ($W=0$) scalar field from $\Lambda$CDM  model by SKA1-mid. Although we show it for linear potential, but the result holds for other potentials as well.

\begin{figure*}
\begin{center} 
\resizebox{200pt}{160pt}{\includegraphics{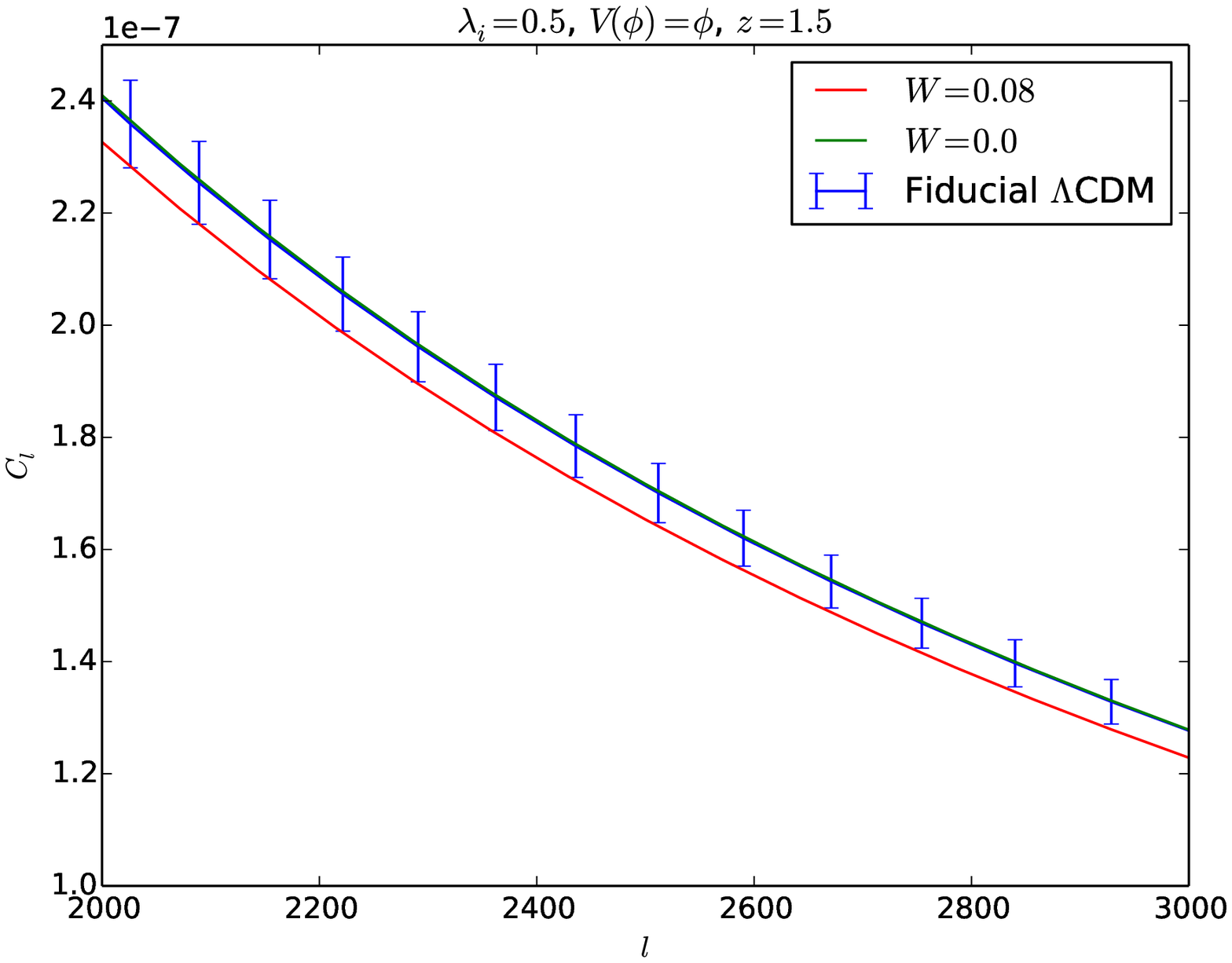}}
\hspace{1mm} \resizebox{200pt}{160pt}{\includegraphics{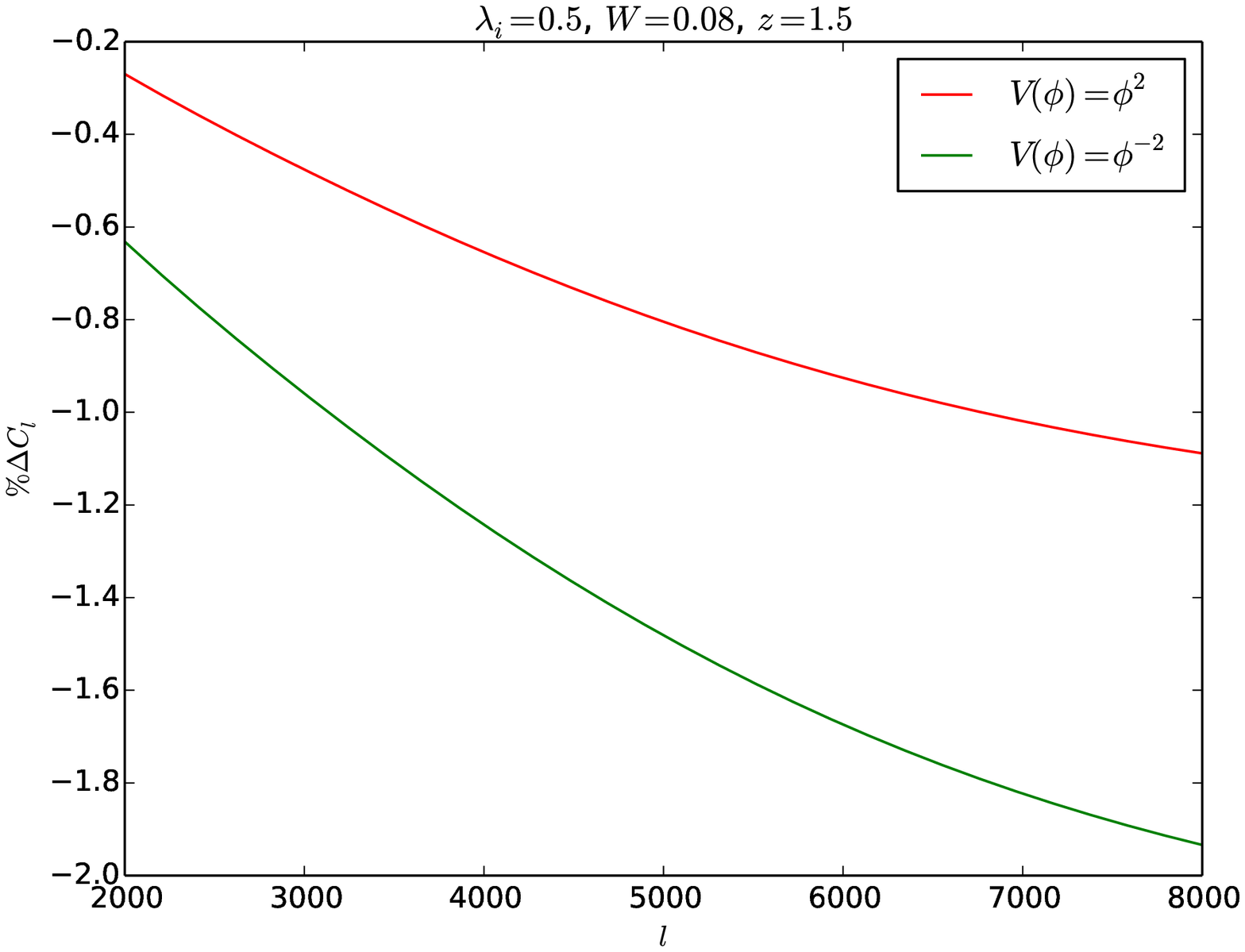}}\\
\end{center}
\caption{Left: Angular power spectra for the fluctuations in HI 21-cm brightness tempertaure for different dark scalar field wih linear potential for coupled and uncoupled case. Right: Percentage deviation in $C_{l}$ for square and inverse-square potentials from the linear potential in coupled case. Here $\lambda_{i} = - \frac{1}{V(\phi)}\frac{d V(\phi) }{d \phi}|_{z=1000}$  }
\end{figure*}

Finally to see whether we can distinguish between different potentials for coupled scalar field models, in figure 2 (right), we show the percentage deviation in $C_{l}$ for square and inverse square potentials from the linear potentials. It shows that inverse-square potential has a greater deviation from linear potential and hence is more probable to be distinguished from the linear potential.

\section{Constraints from BAO measurements using different diagnostics}

Primordial cosmological density fluctuations drive acoustic waves in
the cosmological baryon-photon plasma which gets frozen once
recombination takes place at about $ z \sim 1000$. This leaves a
distinct oscillatory signature on the CMBR temperature anisotropy
power spectrum \citep{peeb70}. The sound horizon at the epoch of
recombination sets a standard ruler that maybe used to calibrate
cosmological distances.  Baryons contribute to $15 \% $ of the total
matter budget, and the baryon acoustic oscillations (BAO) are hence
also imprinted in the late time clustering of non-relativistic
matter. The baryon acoustic oscillation (BAO) is a powerful probe of
cosmological parameters \citep{seoeisen}. Measuring BAO signal
allows us to measure the angular diameter distance $D_A$ and the Hubble
parameter $H(z)$ as functions of redshift using the the transverse and the
longitudinal oscillations respectively. These provide means for
 placing strong  constraints on dark energy models. 
The cross-correlation of the Ly-$\alpha$ forest and redshifted 21-cm
emission  from the post-reionization epoch is believed to be a tool for
mapping out the large-scale structures for redshifts  $z
\le 6$. This mitigates several observational issues like foreground subtraction which poses serious challenge towards detecting the 21-cm signal. The possibility of detecting the BAO signal with the cross-correlation signal is investigated in
\citep{guha13}.

For the Ly-$\alpha$ forest survey, we have considered typical surveys
like BOSS and BIGBOSS which are expected have a high number density of
quasars of $16 \ {\rm deg}^{-2}$ and $64 \ {\rm deg}^{-2}$
respectively. 
We are interested in measuring $\delta q_1 = \delta H(z)/  H(z)$, 
$\delta q_2 = \delta D_A / D_A$ and also $ \delta D_V / D_V$  where   $D_V^3 =  (1 + z)^2 D_A(z) c z/ H(z)$  is a combined distance measure
using the cross-correlation power spectrum $P_{ FT }$ between the Lyman-slpha transmitted flux and the 21-cm brightness temperature. We estimate the errors  using a Fisher matrix  analysis.
The cross-power spectrum is given by 
\begin{equation}
P_{FT} ({\bf k}) = \bar{ T}(z){\bar x}_{\rm HI} b_T C_F ( 1 + \beta_F \mu^2 ) (  1 + \beta_T \mu^2 )P(k)
\end{equation}
We have adopted $(C_F,\beta_F)= (=0.15, 1.11) $ from numerical simulation of the Lyman-alpha forest \citep{mcd03}.

The $ 2 \times 2$ Fisher matrix for estimation of parameters is given by 
\begin{equation}
F_{ij} = \frac{\cal V}{2 \pi^3} \int \frac{d^3{\bf k}}{\left[ P_{FT}^2  + (P_F + N_F)(P_T + N_T)\right] }  \frac{\partial  P_{FT}}{\partial q_i}  \frac{\partial  P_{FT}}{\partial q_j} 
\end{equation}
where $\cal V$ is the survey volume and $P_F$ and $P_T$ denote the auto power spectrum of Lyman alpha forest and  21 cm signal respectively, and $N_F$ and $N_T$ denote the corresponding noise power spectra.
We have $N_F = \frac{1}{\bar n_Q}  (P_{F}^{1D} + \sigma^2 ) $, where $\bar n_Q$ is the quasar density in the field. The aliasing term $ P_{F}^{1D}$ is obtained by integrating the 3D power spectrum and the pixel noise contribution $\sigma^2$ is obtained by assuming that the spectra  are  measured at an average $3-\sigma$ level.
The details of the analysis maybe found in \citet{TGS15}. The Cramer-Rao bound on the error are given by  $\delta q_i = \sqrt {F^{-1}_{ii}} $ and $ \delta D_V / D_V = \frac{1}{3} ( 4 F^{-1}_{11} +  4 F^{-1}_{12} +   F^{-1}_{22} )^{0.5}$.
The noise power spectrum for the 21 cm signal is given by 
\begin{equation}
N_T = 1.0 \times 10^{-3} mK^2 \left( \frac{500}{N_{ant}} \right)^2 \left( \frac{100\rm KHz}{\Delta \nu } \right) \left( \frac{1000 hrs}{t_{obs}}\right). \end{equation}
where $N_{ant}$, $ \Delta \nu$ and $t_{obs}$ denote number of antennae, bandwidth and observation time per pointing. Here we have assumed that two polarizations have been used. 
We have used a $32$MHz bandwidth observation and the same number of antennae as before. For simplicity, here we have assumed antenna distribution to be such that it  gives an uniform distribution of baselines.
We  find that it is necessary to
achieve a noise level of $N_T = 1.1 \times 10^{-5} \ {\rm mK}^2$ and $N_T = 6.25
\times 10^{-6} \ {\rm mK}^2$ per field of view in the redshifted 21-cm
observations to detect the angular and radial BAO respectively with
BOSS at a redshift $z=2.5$.  The corresponding noise levels are $3.3 \times 10^{-5} \ {\rm mK}^2$
and $1.7 \times 10^{-5} \ {\rm mK}^2$ for BIGBOSS. These noise levels should be
attainable by $10000$ hrs observations distributed equally over  100 pointings with SKA1-mid roughly covering the 
full sky coverage of BOSS. It is possible to measure the distance measure $D_V$ 
with $D_V^3 =  (1 + z)^2 D_A(z) c z/ H(z)$ at $\delta D_V/ D_V = 2 \%$  from the cross-correlation signal with $N_T = 3 \times 10^{-4}$ mK{$^2$}. We have
$(\delta D_V/ D_V, \delta D_A/ D_A, \delta H/H) = ( 1.3, 1.5, 1.3 )\% $  and $
(0.67, 0.78, 0.74)\%$ at $N_T =10^{-4}{\rm mK}^2$ and $N_T= 10^{-5}{\rm mK}^2$  at the fiducial redshift $z = 2.5$. The quasar pixel noise is held constant to a $3-\sigma$ in units of the mean transmitted flux. The results are shown in figure 3.

\begin{figure*}
\begin{center} 
\resizebox{200pt}{160pt}{\includegraphics{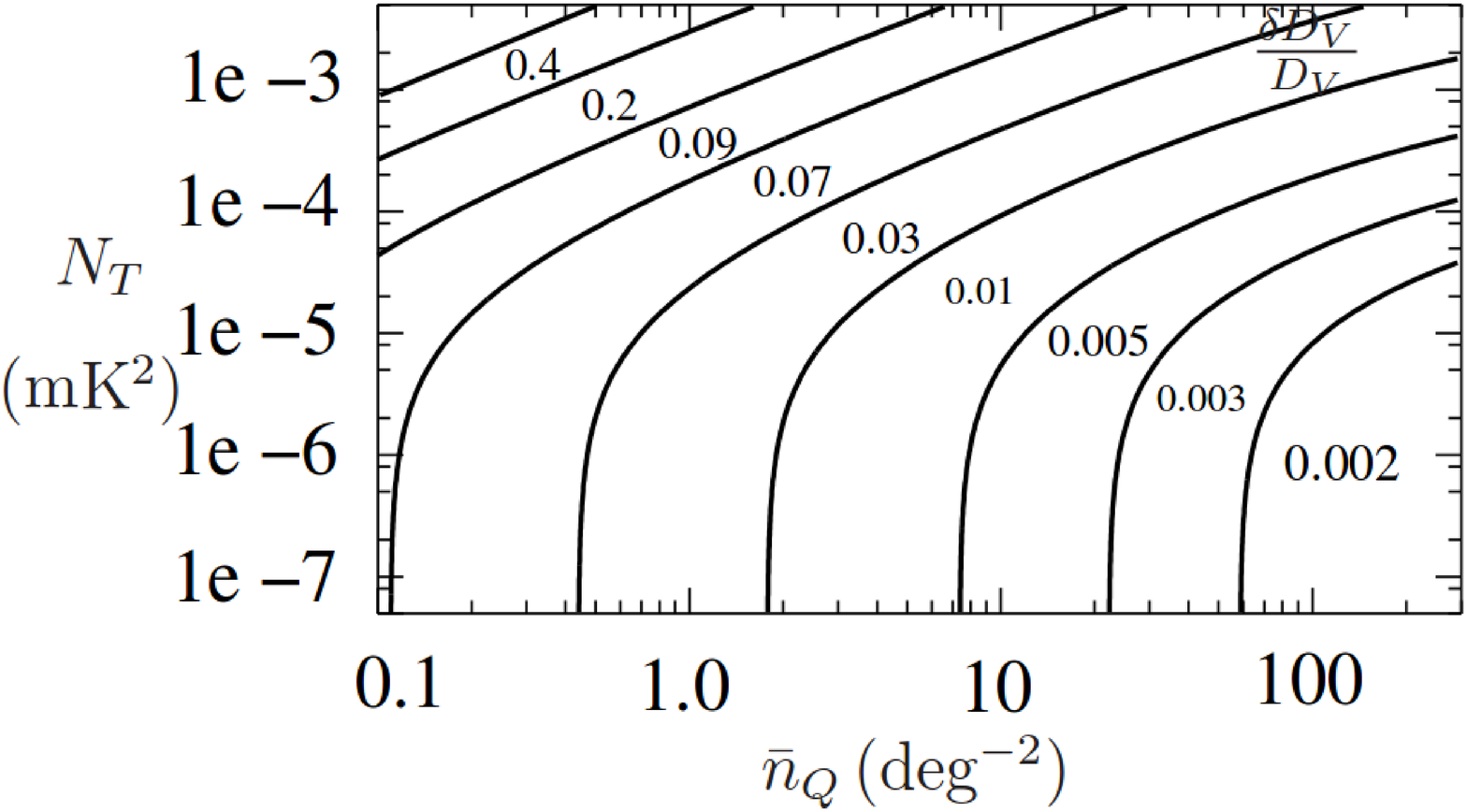}}
\hspace{1mm} \resizebox{200pt}{160pt}{\includegraphics{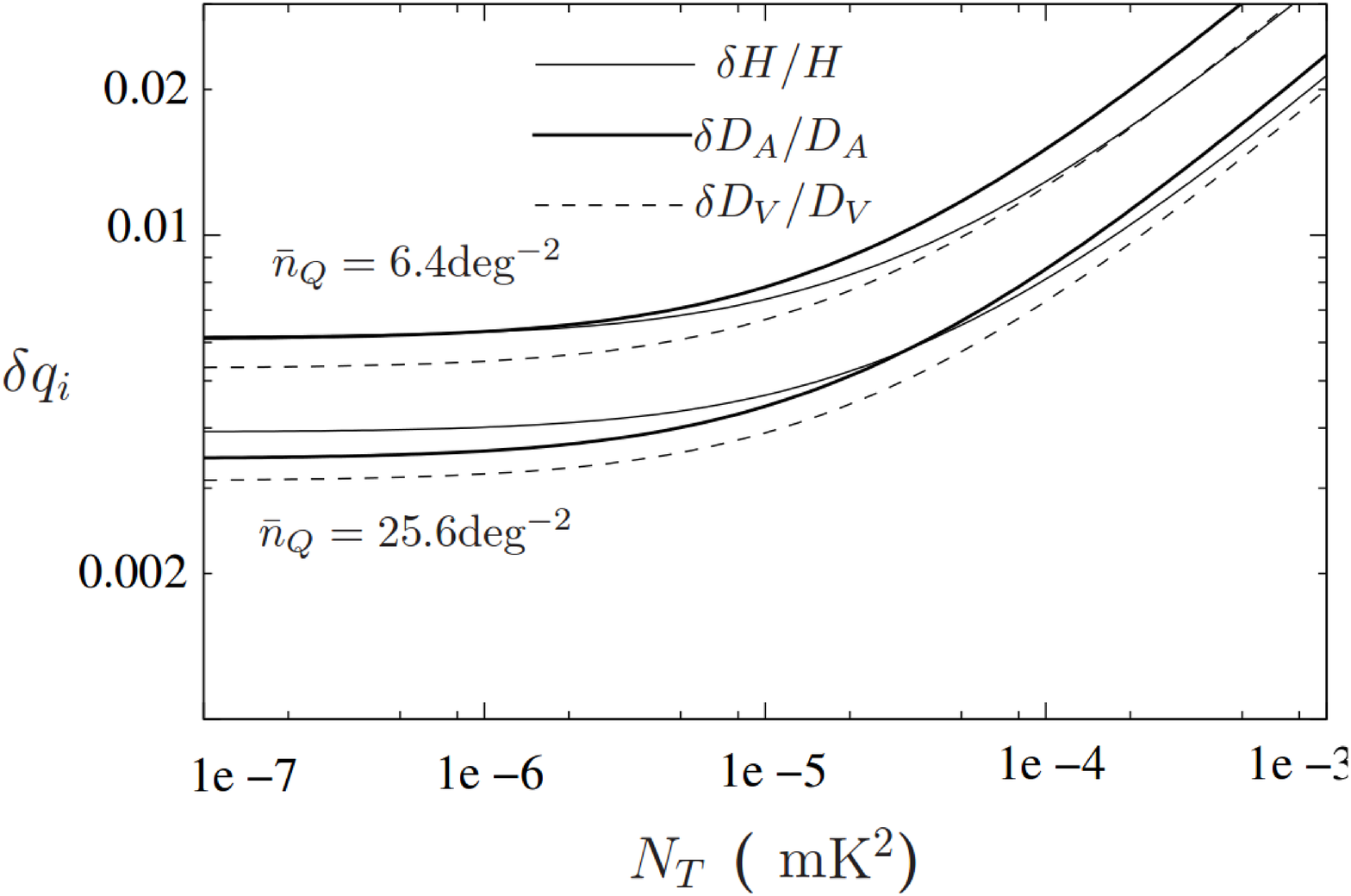}}\\
\end{center}
\caption{Left: The error contours for the distance measure $D_V$ Right: Error prediction for BOSS and BIGBOSS. Figures from \citep{guha13}. We find that at large $N_T$ there is a rough scaling $\delta q_i \propto N_T^ {0.5}$. } 
\end{figure*}

%\begin{figure}
%\begin{center}
% \mbox{\epsfig{file=fig1.eps,width=0.65505\textwidth,angle=0}}
%\caption{The error contours for the distance measure. }
%\label{fig:kappa1}
%\end{center}
%\end{figure}
%\begin{figure}
%\begin{center}
%\mbox{\epsfig{file=fig2.eps,width=0.65505\textwidth,angle=0}}
%\caption{Error prediction for BOSS and BIGBOSS}
%\label{fig:kappa1}
%\end{center}
%\end{figure}

\section{Summary}

The 21~cm line from HI in the post-reionization universe can be a strong probe of cosmology and galaxy formation. We discuss the efforts of the Indian scientists in this area in the context of future SKA survey. We discuss the predictions and forecasts for probing  a large class of scalar field dark energy models using  angular  power  spectra for  the intensity mapping from SKA. Although we restrict ourselves to canonical scalar field models, but the study can be easily extended to more general class of non-canonical and Hordenski class of models for scalar field. We also show that the intensity mapping experiments can be useful in probing the BAO, in particular, when the data is cross correlated with Ly$\alpha$ forest experiments like the BIGBOSS. And this in turn can be used to constrain dark energy models.

The main obstacle in detecting the cosmological redshifted 21-cm signal is the presence of foregrounds from astrophysical galactic and extra galactic sources. These are several orders of magnitude larger than the actual signal. We should therefore carefully subtract these foregrounds before an accurate statistical detection of the cosmological signal is confirmed. In this regard, the cross-correlation studies discussed here are expected to be more efficient in dealing with the experimental systematics. With a number of facilities (SKA, Euclid, LSST) expected to be active in different wave bands in coming years, the future thus looks promising.

\bibliographystyle{mnras}
\bibliography{ska_rev}

\end{document}